\documentclass[natbib,graybox]{svmult}

\usepackage{mathptmx}       
\usepackage{helvet}         
\usepackage{courier}        
\usepackage{type1cm}        
\usepackage{makeidx}         
\usepackage{graphicx}        
\usepackage{multicol}        
\usepackage[bottom]{footmisc}

\usepackage{amsmath}
\usepackage{amssymb}

\makeindex             
\newcommand{\Order}{\mbox{$\mathcal{O}$}} 
\newcommand\Alfven{Alfv\'en }
\newcommand\Alfvenic{Alfv\'enic }
\newcommand{\V}[1]{\mathbf{#1}}

\newcommand{\figref}[1]{Figure~\ref{#1}}

\newcommand{\secref}[1]{\S\ref{#1}}
\renewcommand{\eqref}[1]{equation~(\ref{#1})}

\begin{document}

\title*{Kinetic Turbulence}
\author{Gregory G. Howes}
\institute{Gregory G. Howes \at Department of Physics and Astronomy, 
University of Iowa, Iowa City, IA 52242, USA. \email{gregory-howes@uiowa.edu}}

%
\maketitle

\abstract{The weak collisionality typical of turbulence in many diffuse 
astrophysical plasmas invalidates an MHD description of the turbulent
dynamics, motivating the development of a more comprehensive theory of
kinetic turbulence. In particular, a kinetic approach is essential for
the investigation of the physical mechanisms responsible for the
dissipation of astrophysical turbulence and the resulting heating of
the plasma. This chapter reviews the limitations of MHD turbulence
theory and explains how kinetic considerations may be incorporated to
obtain a kinetic theory for astrophysical plasma turbulence. Key
questions about the nature of kinetic turbulence that drive current
research efforts are identified.  A comprehensive model of the kinetic
turbulent cascade is presented, with a detailed discussion of each
component of the model and a review of supporting and conflicting
theoretical, numerical, and observational evidence.}

\section{Introduction}
The study of turbulence in astrophysical plasmas has almost
exclusively employed a magnetohydrodynamic (MHD) description of the
turbulent dynamics, treating the magnetized plasma as a single fluid,
an approximation valid for large-scale, low-frequency dynamics in the
strongly collisional limit. Yet, the plasmas in a wide variety of
turbulent astrophysical environments often violate one or more of the
conditions required by the MHD approximation, particularly on the
small scales at which dissipation mechanisms act to damp the turbulent
fluctuations, ultimately leading to heating of the plasma. The study
of the turbulent dynamics at small scales and of the physical
mechanisms responsible for the dissipation of the turbulence generally
requires a kinetic treatment.  Thus, it is necessary to leave behind
the comfortable surroundings of the theory of MHD turbulence and enter
the uncharted territory of the evolving theory of
\emph{kinetic turbulence}.

\subsection{Quantitative Characterization of  Plasma Turbulence}

Turbulent systems are typically described theoretically by a spectral
decomposition of the broadband spatial fluctuations into a sum of
plane wave modes, each characterized by its three dimensional
wavevector, phase, and amplitude. An energy spectrum of the turbulent
fluctuations therefore provides a useful quantitative description of
the turbulent system. In a magnetized plasma, the three-dimensional
wavevector space can be reduced to two dimensions by assuming axial
symmetry about the direction of the equilibrium magnetic field,
requiring only the specification of the turbulent power with respect
to the cylindrical components of the wavevector, $k_\perp$ and
$k_\parallel$.  The nature of the dynamics in the different ranges of
the kinetic turbulent cascade can be quantitatively characterized by
two properties: (1) the
\emph{one-dimensional magnetic energy spectrum} in perpendicular
wavenumber, $E_B(k_\perp)$; and (2) the
\emph{wavevector anisotropy}, or the distribution of turbulent power
in wavevector space.  Here $E_B(k_\perp)$ is defined such that the
total magnetic energy is given by $E_B = \int dk_\perp
E_B(k_\perp)$. For \Alfvenic turbulence that is driven isotropically
at the outer-scale wavenumber $k_0$, the conjecture of critical balance
implies that the turbulent power fills a region of the cylindrical
wavevector space satisfying $k_\parallel \lesssim k_0^{1-q}
k_\perp^q$. Specification of the scaling of the boundary of this
region, $k_\parallel \propto k_\perp^q$, is sufficient to completely
characterize the anisotropic distribution of turbulent power.

\subsection{Limits of MHD Treatment of Astrophysical Turbulence}
The limitations of an MHD treatment of astrophysical turbulence can be
illuminated by considering the domain of applicability of MHD
turbulence theory within the broader context of plasma turbulence.
Beginning with the general theory of the turbulent cascade of kinetic
energy in hydrodynamic systems, we consider the modifications required
to describe the turbulent energy cascade in the magnetized plasma
systems relevant to astrophysical environments.

\subsubsection{From Fluid to Kinetic  Models of the Turbulent Cascade}
\label{sec:models}
The limitations of MHD turbulence theory can be illustrated most
clearly by a qualitative comparison of the features of nonlinear
cascade of energy in hydrodynamic turbulence, MHD turbulence, and
kinetic turbulence.

\begin{figure}[top] 
\vspace*{2mm}
\begin{center}
\resizebox{10.cm}{!}{
\includegraphics*[1.35in,2.85in][7.8in,7.7in]{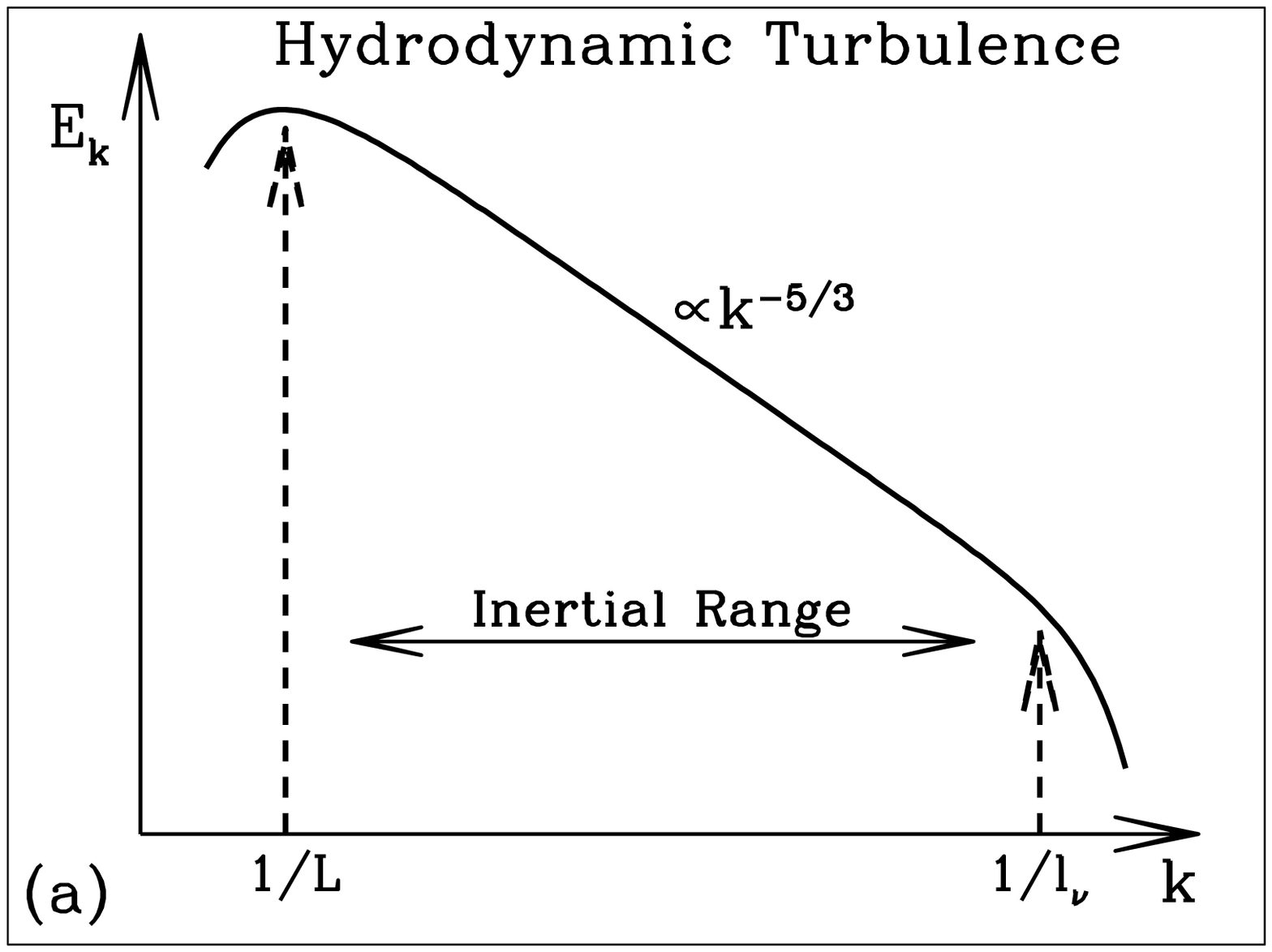}} 
\resizebox{10.cm}{!}{
\includegraphics*[1.35in,2.85in][7.8in,7.7in]{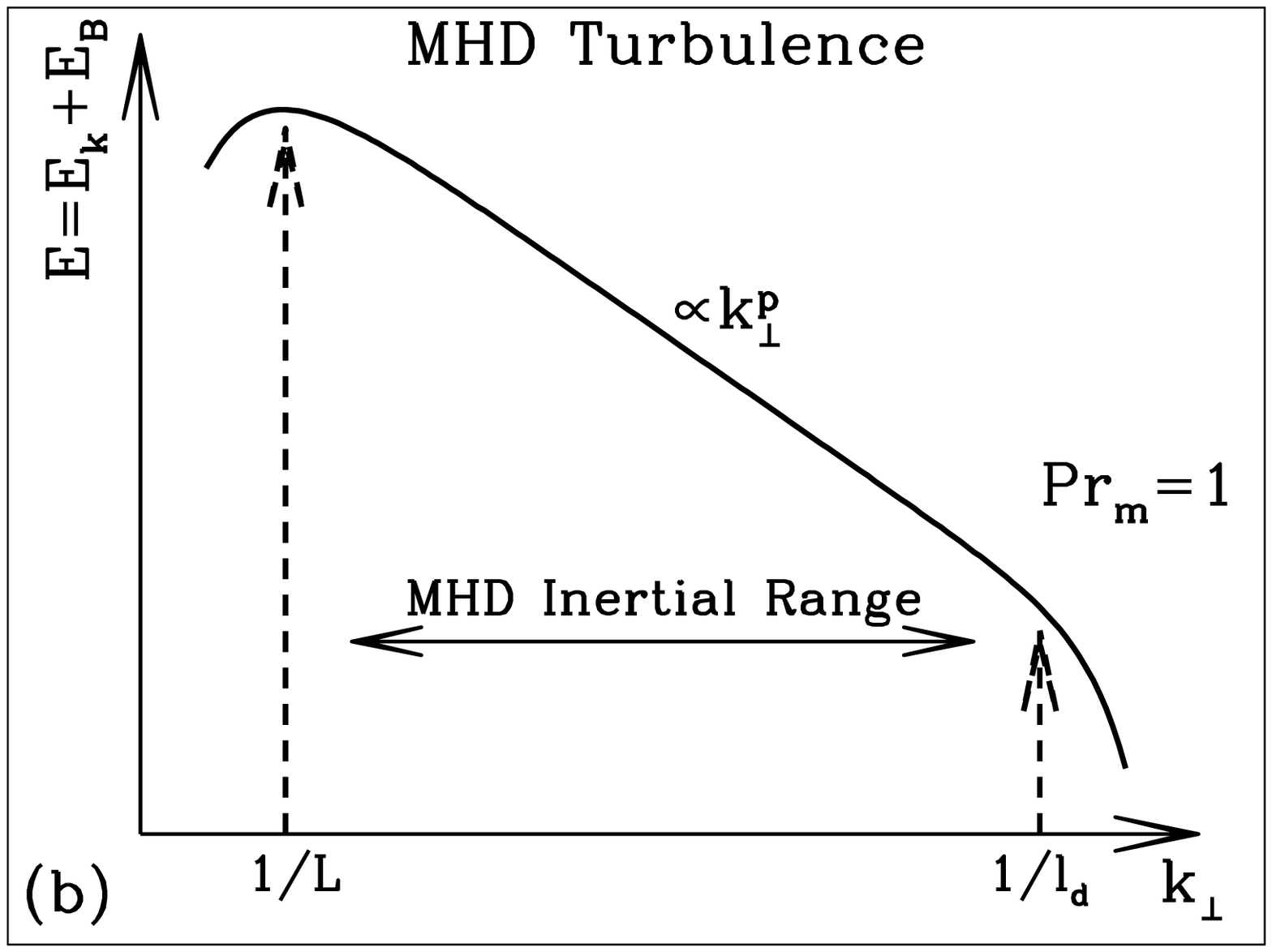}}
\end{center}
\caption{(a) Wavenumber spectrum for kinetic energy in hydrodynamic turbulence, 
from the driving scale, $L$, through the inertial range, to the viscous
dissipation scale, $l_\nu$. (b) Perpendicular wavenumber spectrum for
total energy $E=E_k+E_B$ in MHD turbulence with $\mbox{Pr}_m=1$, from
the driving scale, $L$, through the inertial range, to the viscous and
resistive dissipation scale, $l_d=l_\nu=l_\eta$.
\label{fig:hd_mhd} }
\end{figure}

In hydrodynamic systems, turbulent motions are driven at some large
scale $L$, denoted the driving or energy injection scale.  Nonlinear
interactions serve to transfer the turbulent kinetic energy to motions
at ever smaller scales, until reaching a small scale $l_\nu$ at which
dissipation via viscous damping is sufficient to terminate the
turbulent cascade.  For typical hydrodynamic systems, a large dynamic
range exists between the driving and dissipation scales, $L/l_\nu \gg
1$. In that case, one may define an \emph{inertial range} of scales
$l$ within which the effects of the driving and dissipation are
negligible, $L\gg l \gg l_\nu$.  Within the inertial range, there
exists no particular characteristic length scale, so the dynamics of
the turbulence in the inertial range is found to be self-similar, and
a simple application of dimensional analysis is sufficient to describe
accurately the steady-state hydrodynamic turbulent cascade of energy
\citep{Kolmogorov:1941}. A qualitative diagram of the kinetic energy
wavenumber spectrum for the hydrodynamic turbulence cascade is shown
in \figref{fig:hd_mhd}(a).

In magnetohydrodynamic systems, the turbulence theory must be modified
in three important ways. First, the dynamics of two turbulent fields,
the velocity and the magnetic field, must be described, so the cascade
of both kinetic and magnetic energy is mediated by nonlinear turbulent
interactions.  Second, fluctuations of the two turbulent fields are
dissipated by distinct mechanisms, viscosity for the velocity and
resistivity for the magnetic field. The characteristic length scales
of viscous dissipation $l_\nu$ and resistive dissipation $l_\eta$ need
not be equal, and their ratio is characterized by the magnetic Prandtl
number $\mbox{Pr}_m \equiv l_\nu/l_\eta$. Third, the magnetic field in
the plasma establishes a preferred direction, leading to distinct
dynamics in the direction parallel to the magnetic field and in the
plane perpendicular to the magnetic field. In addition, the magnetic
tension provided by the magnetic field supports a type of linear wave,
the \Alfven wave, which has no counterpart in the hydrodynamic case,
transforming the nature of the turbulent motions from hydrodynamic
vortices to magnetohydrodynamic waves.  This third complication is the
most significant change from hydrodynamic turbulence, and leads to the
inherent anisotropy of MHD turbulence, where turbulent energy is
transferred more rapidly to small perpendicular scales $l_\perp$ than
to small parallel scales $l_\parallel$. Nonetheless, despite these
significant differences, the overall qualitative picture of the
turbulent energy cascade in MHD turbulence bears a striking
resemblance to the hydrodynamic case.

Consider, in particular, the simplified case of MHD turbulence in a
$\mbox{Pr}_m =1$ plasma, so there exists a single dissipation scale
$l_d=l_\nu=l_\eta$.  One may define an MHD inertial range, $L\gg
l_\perp \gg l_d$, directly analogous to the hydrodynamic case.  Due
the anisotropy of the turbulent energy transfer, the turbulent
dynamics are optimally described with respect to the perpendicular
scale $l_\perp$. The evolution of the parallel scale is determined in
terms of the perpendicular scale by the condition of \emph{critical balance}
\citep{Goldreich:1995}, so that $l_\parallel  \propto l_\perp^{q}$.
The exponent $q$ describes the \emph{scale-dependent anisotropy} of the MHD
turbulent cascade, where $q=2/3$ in the Goldreich-Sridhar model
\citep{Goldreich:1995}, and $q=1/2$ in the Boldyrev model
\citep{Boldyrev:2006}. Similar to the hydrodynamic case, in the
MHD inertial range, there exists no characteristic length scale, so
the dynamics of MHD turbulence is found to be self-similar as
well. Therefore, for the $\mbox{Pr}_m =1$ case, the MHD turbulence
theory appears nearly the same as the hydrodynamic turbulence theory,
with a few minor changes: (1) the turbulent cascade is described by
the perpendicular scale $l_\perp$ rather than an isotropic scale $l$;
(2) there exists a scale-dependent anisotropy due to the parallel
scaling $l_\parallel \propto l_\perp^{q}$; and (3) the exponent $p$ in
the self-similar power law solution for the one-dimensional energy
spectrum $E \propto k_\perp^p$ may differ quantitatively from the
hydrodynamic solution. But the general qualitative picture---a
self-similar MHD turbulent cascade of energy from the driving scale
$L$, through an inertial range, to the dissipative scale
$l_d$---remains essentially the same as the hydrodynamic cascade, as
is evident by comparing the diagram of the wavenumber spectrum for
total energy $E=E_k+E_B$ in the MHD turbulent cascade in
\figref{fig:hd_mhd}(b) to the hydrodynamic case in
\figref{fig:hd_mhd}(a).

In kinetic plasma systems, this simple qualitative model of the
turbulent cascade changes dramatically due to the existence of three
characteristic length scales and new physics associated with each of
these scales. The three characteristic length scales that come into
play in typical conditions for turbulent astrophysical plasmas are the
ion mean free path $\lambda_i$, the ion Larmor radius $\rho_i$, and
the electron Larmor radius $\rho_e$. The MHD approximation requires
the following four conditions:
\begin{enumerate} 
\item  Nonrelativistic conditions, $v_{ts}/c
\ll 1$
\item  Strongly collisional conditions, $\lambda_i/l
\ll 1$
 \item Large-scale motions, $\rho_i/l \ll 1$
\item Low-frequency dynamics, $\omega/ \Omega_i \ll 1$
\end{enumerate}
Here $v_{ts}=\sqrt{2 T_s/m_s}$ is the thermal velocity\footnote{Here
$T_s$ is expressed in units of energy, absorbing the Boltzmann
constant.} of species $s$, $\omega$ is the typical frequency of the
turbulent fluctuations, and $\Omega_i$ is the ion cyclotron
frequency. It is clear that, in the MHD approximation, all three of
the characteristic scales above are assumed to be infinitesimal
compared to the typical scale of the turbulent motions, $l$. However,
in astrophysical plasmas of interest, the turbulent dynamics
frequently violate conditions (2) and (3) above\footnote{Note that
condition (4) is not generally independent of condition (3).  For MHD
\Alfven waves, the condition $\omega \ll \Omega_i$ may be
alternatively written $ \rho_i/l_\parallel \ll \sqrt{\beta_i}$, where
the ion plasma beta is $\beta_i= 8 \pi n_i T_i/B^2$. Thus, if
$\sqrt{\beta_i} \sim \Order(1)$, then condition (4) is roughly
equivalent to condition (3).}. Therefore, it is important to examine
more closely how these characteristic scales enter into the dynamics
of the turbulent cascade in astrophysical plasmas, leading to a
violation of the MHD approximation and requiring the transition to a
kinetic description of the turbulent dynamics.

\subsubsection{Violation of the MHD Approximation}

\begin{figure}[top] 
\vspace*{2mm}
\begin{center}
\resizebox{10.cm}{!}{
\includegraphics*[1.35in,2.85in][7.8in,7.7in]{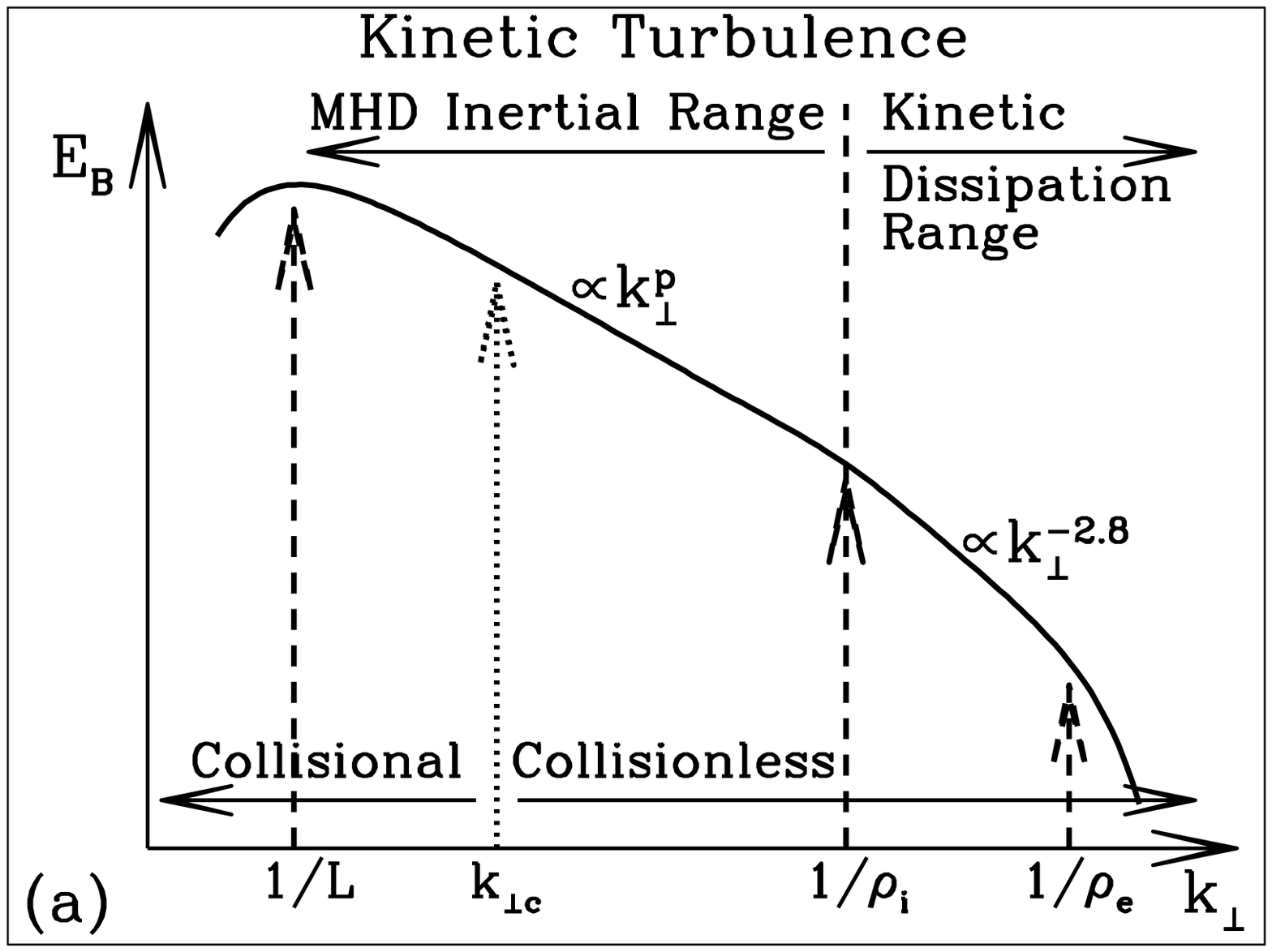}} 
\resizebox{10.cm}{!}{
\includegraphics*[1.35in,2.85in][7.8in,7.7in]{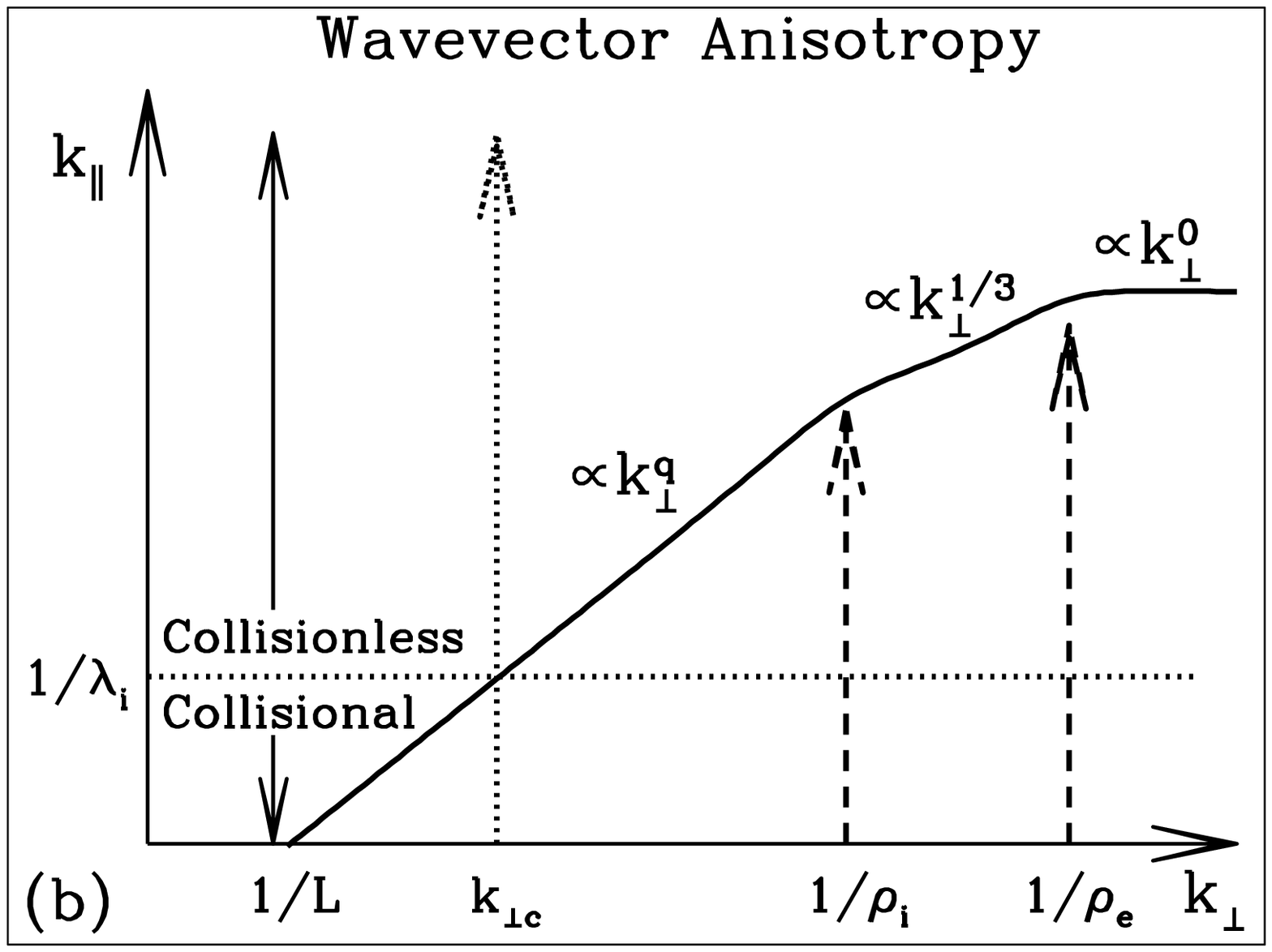}}
\end{center}
\caption{(a) Perpendicular wavenumber spectrum for magnetic energy in kinetic turbulence,
from the driving scale, $L$, through the MHD inertial range to the ion
Larmor radius $\rho_i$, where the turbulent cascade enters the kinetic
dissipation range, and down to the electron Larmor radius
$\rho_e$. The transition from collisional to collisionless dynamics
occurs at $k_{\perp c}$. (b) Wavevector anisotropy in kinetic
turbulence, scaling as $k_\perp^q$ in the MHD inertial range,
$k_\perp^{1/3}$ in the kinetic dissipation range, and $k_\perp^{0}$
(no parallel cascade) beyond electron scales. The transition from
collisional to collisionless dynamics occurs at $k_{\parallel} \rho_i
\sim 1$.
\label{fig:kinetic} }
\end{figure}

Spacecraft measurements of turbulence in the solar wind provide
invaluable guidance for the construction of a theoretical model that
describes the energy spectrum of the kinetic turbulent cascade.
Recent measurements of solar wind turbulence with unprecedented
temporal resolution enable us to probe the turbulent dynamics down to
the scale of the electron Larmor radius
\citep{Sahraoui:2009,Kiyani:2009,Alexandrova:2009,Chen:2010b,Sahraoui:2010b,Alexandrova:2012}. Therefore, we now have a fairly complete 
observational picture of the kinetic turbulent cascade in the solar
wind over a dynamic range of $10^6$ from the the large energy
injection scale at $L\sim 10^6$~km down to the scale of the electron
Larmor radius at $\rho_e \sim 1$~km. From the large body of turbulence
measurements in the solar wind
\citep{Sahraoui:2009,Kiyani:2009,Alexandrova:2009,Chen:2010b,Sahraoui:2010b,Alexandrova:2012},
we can construct a general diagram for the perpendicular wavenumber
spectrum of the magnetic energy in turbulent astrophysical plasmas,
shown in \figref{fig:kinetic}(a). It is important to emphasize here that, although
the general form of the magnetic energy spectrum is well established
from observations, the interpretation of this spectrum in terms of the
characteristic plasma scales requires significant input from plasma
kinetic theory, and many of the features of \figref{fig:kinetic}(a) remain topics of
active research.

In \figref{fig:kinetic}(a), the plasma turbulence is driven at some large scale $L
\gg \rho_i$. It is generally assumed, in the absence of arguments to
the contrary, that the turbulence is driven isotropically with respect
to the magnetic field, so that the perpendicular and parallel
components of the driving wavevector $\V{k}_0$ are equal,
$k_{\parallel 0} \sim k_{\perp 0} \sim k_0 \sim 1/L$. If the plasma
conditions at the driving scale satisfy the MHD approximation, then
the large scale end of the turbulent cascade is described by MHD
turbulence theory. Although the turbulent fluctuations in an MHD
plasma may, in general, be composed of a mixture fast, Alfv\'en, and
slow waves, observational and numerical evidence suggests that \Alfven
waves dominate the turbulent dynamics in typical astrophysical plasmas
(this point is discussed further below). For the \Alfvenic turbulent
cascade, the one-dimensional magnetic energy spectrum as a function of
perpendicular wavenumber $k_\perp$ scales as $E_B \propto k_\perp ^p$,
where the spectral index is $p=-5/3$ in the Goldreich-Sridhar model
\citep{Goldreich:1995} or  $p=-3/2$ in the Boldyrev model
\citep{Boldyrev:2006}. The wavevector anisotropy of the anisotropic 
\Alfvenic  cascade scales as $k_\parallel\propto k_\perp^q$, 
where the values for $q$ are given in \secref{sec:models}; this
anisotropic cascade of energy through wavevector space is depicted in
\figref{fig:kinetic}(b).

As the MHD turbulent cascade transfers energy to smaller scales
(higher wavenumber), it eventually reaches the one of the
characteristic length scales $\lambda_i$, $\rho_i$, or $\rho_e$, at
which point the MHD approximation is violated.  Here we focus on
exploring how these length scales enter into the model for kinetic
turbulence and what effect they have on the turbulent dynamics. For
typical conditions in astrophysical plasmas, the characteristic length
scales are ordered by $\lambda_i > \rho_i > \rho_e$, so the ion mean
free path $\lambda_i$ is usually reached first.

The ion mean free path $\lambda_i$ characterizes the collisionality
for the motion of plasma particles parallel to the magnetic
field\footnote{The Lorentz force limits the perpendicular motion of
plasma particles to the particle Larmor radius. Since typical
astrophysical conditions yield $\rho_i \ll \lambda_i$, the plasma is
essentially always collisionless in the perpendicular direction. Note,
however, that because plasma particles cannot move beyond the Larmor
radius in the perpendicular direction from the magnetic field, this
embodies the large-scale \emph{perpendicular} motions, $l_\perp \gg
\rho_i$, with a fluid-like behavior, even under weakly collisional
conditions.}, so it must be compared to the parallel wavenumber
$k_\parallel$.  For $k_\parallel \lambda_i \ll 1$, the plasma is
strongly collisional; for $k_\parallel
\lambda_i \gg 1$, the plasma is weakly collisional. Fluid
approximations, such as hydrodynamics or MHD, break down for plasma
conditions $k_\parallel \lambda_i \gtrsim 1$, so kinetic theory is
formally required to describe the plasma dynamics in moderately to
weakly collisional regimes.

As depicted in \figref{fig:kinetic}(b), at some point in the MHD inertial range, the
parallel scales may reach the scale of the ion mean free path,
$k_\parallel \lambda_i \sim 1$, marking the transition from collisional dynamics 
that is well described by MHD at $k_\parallel \lambda_i \ll 1$ to
collisionless dynamics that requires a kinetic description at $k_\parallel
\lambda_i \gg 1$.  The condition of critical balance determines the 
relation between the parallel and perpendicular wavenumbers of strong
MHD turbulence \citep{Goldreich:1995}, so we may define the
perpendicular wavenumber $k_{\perp c}$ that corresponds to the
transition of collisionality at $k_\parallel \lambda_i \sim 1$.  This
transition in the perpendicular wavenumber spectrum typically occurs
at perpendicular scales larger than the ion Larmor radius, $k_{\perp c
} \rho_i < 1$, as shown in \figref{fig:kinetic}(a).  For perpendicular wavenumbers
$k_{\perp} \ll k_{\perp c } $, the strongly collisional dynamics is
well described by MHD, and for $k_{\perp} \gg k_{\perp c } $ the
weakly collisional dynamics require a kinetic description, as depicted in 
\figref{fig:kinetic}(a).

For the weakly collisional range $k_{\perp} \gg k_{\perp c }$, it has
been shown rigorously from kinetic theory that the \Alfvenic turbulent
fluctuations remain essentially fluid in nature
\citep{Schekochihin:2009}. The  \Alfvenic turbulent cascade continues to be 
accurately described by the equations of reduced MHD
\citep{Strauss:1976} and remains undamped down to the perpendicular
scale of the ion Larmor radius, $k_\perp \rho_i \sim 1$
\citep{Schekochihin:2009}. Therefore, although the MHD approximation 
is formally violated at scales $k_{\perp} \gtrsim k_{\perp c }$, the
MHD description of the anisotropic \Alfvenic cascade remains applicable,
regardless of whether the dynamics is collisional or collisionless,
for all scales larger than the ion Larmor radius, $k_\perp \rho_i \ll
1$.  Therefore, we denote the range of scales
$L \gg l_\perp \gg \rho_i$ in the \emph{kinetic} turbulent cascade as the
\emph{MHD inertial range}.
\begin{svgraybox}
{\bf MHD Inertial Range}: The range of perpendicular scales from the large
scale of energy injection to the scale of the ion Larmor radius, $L
\gg l_\perp \gg \rho_i$, including both collisional and collisionless
regimes.
\end{svgraybox}

On the other hand, compressible turbulent fluctuations associated with
the MHD fast and slow waves in the MHD inertial range require a
kinetic description at all moderately to weakly collisional scales,
$k_\parallel \lambda_i \gtrsim 1$ or $k_{\perp}\gtrsim k_{\perp
c}$. These modes are damped both collisionally by ion viscosity at
$k_\parallel \lambda_i \sim 1$ \citep{Braginskii:1965} and
collisionlessly by ion Landau damping at $k_\parallel \lambda_i \gg 1$
\citep{Barnes:1966}. Therefore, it is expected that the damped
compressible modes will play at most a subdominant role relative to
the undamped \Alfvenic fluctuations in turbulent astrophysical
plasmas.  In the weakly collisional solar wind, for example,
compressible fluctuations generally contribute less than 10\% of the
turbulent magnetic energy
\citep{Tu:1995,Bruno:2005} in the MHD inertial range. Therefore,
 we turn our attention back to the dynamics of the
dominant \Alfvenic turbulent fluctuations.

When the \Alfvenic turbulent cascade reaches the perpendicular scale of
the ion Larmor radius, $k_\perp \rho_i \sim 1$, the MHD description of
the \Alfvenic fluctuations breaks down completely for two reasons.
First, finite Larmor radius effects lead to a decoupling of the ions
from the turbulent electromagnetic fluctuations at perpendicular
wavenumbers $k_\perp \rho_i \gtrsim 1$. The result is that the
non-dispersive \Alfven wave in the limit $k_\perp \rho_i \ll 1$
transitions to the dispersive kinetic \Alfven wave in the limit
$k_\perp \rho_i \gg 1$. The dispersive nature of the \Alfvenic
fluctuations accelerates the rate of the turbulent nonlinear energy
transfer, leading to a steepening of the magnetic energy spectrum,
with a break in the spectrum at the $k_\perp \rho_i
\sim 1$, as shown in \figref{fig:kinetic}(a). Second, collisionless damping of the 
electromagnetic fluctuations occurs due to the Landau resonance with
the ions, with a peak in the ion damping rate around $k_\perp \rho_i
\sim 1$. In addition, electron Landau damping can also contribute
significantly for all scales $k_\perp \rho_i \gtrsim 1$. The combined
effect of the ion and electron collisionless damping can lead to a
further steepening of the spectrum for scales $k_\perp \rho_i \gtrsim
1$ \citep{Howes:2011a,Howes:2011b}. Finally, the cascade reaches the perpendicular
scale of the electron Larmor radius, $k_\perp \rho_e \sim 1$, where
collisionless damping becomes sufficiently strong to terminate the
turbulent cascade, leading to an exponential drop off of the magnetic
energy spectrum \citep{Terry:2012,Alexandrova:2012,TenBarge:2013b}.
MHD turbulence theory cannot describe the dispersive wave behavior or
the dissipation that occurs via kinetic mechanisms at  scales
$k_\perp \rho_i \gtrsim 1$.  Therefore, we denote the range of scales
$l_\perp \lesssim\rho_i$ in the kinetic turbulent cascade as the
\emph{kinetic dissipation range}.

\begin{svgraybox}
{\bf Kinetic Dissipation Range}: The range of perpendicular scales at or
below the scale of the ion Larmor radius, $l_\perp \lesssim\rho_i$,
where wave dispersion and collisionless dissipation play important
roles.
\end{svgraybox}

\subsection{Importance of Kinetic Turbulence}

Of fundamental importance in the study of astrophysical turbulence is
to determine the pathway by which the energy of turbulent motions is
ultimately converted to plasma heat. Astrophysical turbulence is
generally driven by violent events or instabilities at large scales,
but fluctuations are dissipated strongly only at scales of order or
smaller than the ion Larmor radius.  A kinetic turbulent cascade
arises to transfer energy via nonlinear couplings from the large
energy injection scales, through the MHD inertial range, down to the
scale of the ion Larmor radius.  The turbulent fluctuations begin to
be damped when the cascade reaches the scale of the ion Larmor radius,
marking the entry into the kinetic dissipation range. Since the
dynamics within the kinetic dissipation range is typically weakly
collisional, the dissipation of the turbulent electromagnetic
fluctuations must be accomplished via collisionless mechanisms
governed by plasma kinetic theory.  The energy thus removed from the
turbulent fluctuations ultimately leads to thermal heating of the
protons, electrons, and minority ions in the plasma. The observational
signature of astrophysical objects depends strongly on the nature of
the plasma heating, so to interpret observational data requires a
detailed characterization of the small-scale, kinetic plasma
turbulence.

For example, as matter in an accretion disk spirals slowly into a
black hole, it converts a tremendous amount of gravitational potential
energy into heat.  Several physical mechanisms contribute to this
process.  First, the magnetorotational instability
\citep{Balbus:1991,Balbus:1998} taps free energy from the differential
rotation of the accretion disk to drive turbulence on the scale height
of the disk, $L \sim H$. The turbulence effectively transports angular
momentum outward in the disk, enabling accretion disk plasma to fall
down the gravitational potential and mediating the conversion of
gravitational potential energy into kinetic and magnetic energy of the
MHD turbulent fluctuations. The high temperatures characteristic of
the plasma in a black hole accretion disk lead to a collisional mean
free path $\lambda_i \sim H$, so the turbulent dynamics is weakly
collisional. A kinetic turbulent cascade is responsible for the
transfer of turbulent energy through the MHD inertial range down to
the scale of the ion Larmor radius, where the turbulent
electromagnetic fluctuations are damped via collisionless mechanisms
in the kinetic dissipation range. An entropy cascade ultimately
mediates the final conversion of this turbulent free energy into
plasma heat. Therefore, the radiation that is emitted from the hot,
magnetized plasma is a strong function of the black hole properties
and of the character of the small-scale plasma fluctuations, where the
plasma heating occurs.  To interpret observational data from the
Chandra X-ray Observatory, for example, one must unravel the details
of the kinetic turbulent cascade.


Developing a mature model of the kinetic turbulent cascade is critical
to understanding the turbulent dynamics of the kinetic dissipation
range, the physical mechanisms responsible for the damping of the
turbulent fluctuations, and the resulting heating of the plasma
species.  The ultimate goal is to develop a predictive capability to
estimate accurately the heating of the protons, electrons, and
minority ions in the plasma based on the plasma parameters and the
characteristics of the turbulent driving.

\section{Key Questions about Kinetic Turbulence}
\label{sec:ques}
The unprecedented availability of high temporal resolution solar wind
turbulence measurements from current spacecraft missions has enabled
the observational characterization of the kinetic turbulent cascade
from the large scales of energy injection down to the scale of the
electron Larmor radius.  This has spurred the heliospheric physics
community to engage actively the topic of the turbulence in the
dissipation range of the solar wind, and has engendered considerable
controversy about a number of significant issues related to the
fundamental character of kinetic turbulence. In particular, the nature
of both the turbulent fluctuations in this regime and the physical
mechanisms responsible for their dissipation remains highly contested
within the scientific community. Four key questions relevant to the
study of the dissipation range of solar wind turbulence are identified
and are discussed at length in a forthcoming
review \citep{Howes:2014d}:
\begin{enumerate}
\item What are the limits of validity of using a 
fluid description of the turbulence in the dissipation range, and
which aspects of the turbulence require a kinetic description?
\item  Are the linear plasma wave properties
relevant to the turbulent fluctuations of the dissipation range? 
\item What are the characteristic  dynamics of the dissipation range fluctuations?
\item What physical mechanisms are responsible for
the dissipation of the turbulent fluctuations and the ultimate
conversion of their energy to plasma heat?
\end{enumerate}

Although these significant questions about the nature of kinetic
turbulence remain controversial, a promising model of the kinetic
turbulent cascade \citep{Howes:2008c,Howes:2008b,Schekochihin:2009,Howes:2011b}
has been developed that appears to be broadly consistent with most
observations of solar wind turbulence. This model involves an
anisotropic cascade of \Alfvenic fluctuations beginning as a cascade
of \Alfven waves in the MHD inertial range and transitioning to a
cascade of kinetic \Alfven waves subject to collisionless damping in
the kinetic dissipation range.  Yet the cascade of energy from large
to small scales described by this kinetic turbulence model may not
explain all of the fluctuations observed in the solar wind. For
example, fluctuations can be generated by the action of kinetic
temperature anisotropy instabilities \citep{Bale:2009} that are driven
by the spherical expansion of the solar wind, an effect beyond the
scope of this model. Plausible arguments exist that suggest some of
these additional effects may coexist peacefully with the kinetic
turbulence, proceeding without being significantly affected by or
significantly affecting the kinetic turbulent cascade. The remainder
of this chapter aims to describe in detail the model of the kinetic
turbulent cascade and to discuss the supporting and conflicting
theoretical, observational, and numerical evidence.

\section{A Model of the Kinetic Turbulent Cascade}
A basic theoretical model of the kinetic turbulent cascade in
astrophysical plasmas has been developed with the aim to describe
completely the flow of energy from the large driving scales of the
turbulence to its ultimate fate as thermal heat of the plasma
\citep{Howes:2008c,Howes:2008b,Schekochihin:2009,Howes:2011b}. We present here a 
brief outline of this model, before delving into a detailed
description of each component of the model and a discussion of
supporting and conflicting evidence.

Violent events or instabilities first drive turbulent fluctuations of
the magnetic field and plasma at some large scale, generating a
mixture of finite amplitude fast, Alfv\'en, and slow waves. If the
fluctuations are driven isotropically with velocities approximately
equal to the \Alfven velocity in the plasma, a cascade of strong
compressible MHD turbulence will mediate the transfer of the turbulent
kinetic and magnetic energy to smaller scales. The fast waves cascade
to smaller scales isotropically, while the critically balanced \Alfven
wave cascade produces an anisotropic distribution of
\Alfven and slow wave fluctuations in this collisional part of the
MHD inertial range. The parallel scales of the turbulent fluctuations
eventually reach the ion collisional mean free path, marking the
transition from strongly to weakly collisional dynamics. The
compressible fast and slow wave fluctuations suffer collisional
damping at the moderately collisional scale of the transition, and
collisionless damping at the smaller, weakly collisional scales. The
incompressible
\Alfvenic fluctuations remain undamped through this transition,  so
the damped fast and slow waves are expected to contribute
subdominantly to the turbulence compared to the
\Alfven waves.  The \Alfven waves continue their cascade undamped  
through the collisionless remainder of the MHD inertial range until
their perpendicular scales reach the ion Larmor radius, marking the
transition to the kinetic dissipation range.

The anisotropic \Alfvenic fluctuations at this transition transfer
energy into a cascade of kinetic \Alfven waves at perpendicular scales
below the ion Larmor radius. In addition, collisionless wave-particle
interactions via the Landau resonance with the ions lead to a peak in
the ion kinetic damping at the ion Larmor radius, dissipating some
fraction of the turbulent electromagnetic fluctuation energy.  The
undamped remainder of the turbulent energy continues as a cascade of
kinetic \Alfven waves to smaller perpendicular scales, forming the
kinetic dissipation range at all scales below the ion Larmor radius.
Throughout this range, electron Landau damping may cause significant
collisionless damping of the turbulent fluctuations, with the strength
of the damping increasing as the perpendicular scale decreases. At the
perpendicular scale of the electron Larmor radius, the electron Landau
damping becomes sufficiently strong to terminate the cascade, leading
to an exponential decay of the turbulent energy spectrum at the
electron scale.

Thermodynamically, the transfer of free energy from the kinetic and
magnetic energy of the turbulent electromagnetic fluctuations to free
energy in velocity space structure of the particle distribution
functions is not equivalent to irreversible thermal heating of the
plasma. Irreversible plasma heating, and the associated increase of
entropy, ultimately requires collisions.  This is accomplished in a weakly
collisional plasma by the ion and electron entropy cascades, dual
cascades in physical and velocity space that drive fluctuations to
small enough  velocity-space scales that arbitrarily weak
collisions are sufficient to achieve irreversibility. This final
process marks the thermodynamic end of the kinetic turbulent cascade,
completing the conversion of large-scale turbulent fluctuation energy
to thermal heat of the plasma.

This model of the kinetic turbulent cascade implies certain answers to
the questions posed in \secref{sec:ques}, so we elucidate those
answers here:
\begin{enumerate}
\item A fluid description is applicable for all  turbulent
fluctuations at scales larger than the collisional transition, and for
the \Alfvenic dynamics at all scales larger than the ion Larmor
radius. The dynamics and kinetic damping of the compressible
fluctuations at all moderately to weakly collisional scales, and of the 
\Alfvenic fluctuations at the scales of the ion Larmor radius and below, 
require a kinetic description.

\item The properties of the turbulent fluctuations at scales sufficiently
below the driving scale are related to the characteristics of the
linear kinetic plasma waves. 

\item The dissipation range fluctuations are kinetic \Alfven waves.

\item Ion and electron Landau damping are the physical mechanisms 
by which the turbulent electromagnetic fluctuations are damped, and
the ion and electron entropy cascades mediate the irreversible
transition of free energy in the particle distribution functions to
thermal heat.
\end{enumerate}
In the following sections, we describe in detail all of the facets of
this model of the kinetic turbulent cascade, providing supporting
theoretical, observational, and numerical evidence and reviewing
findings in conflict with this model. A general diagram of the
magnetic energy spectrum and the distribution of turbulent power in
wavevector space is shown in \figref{fig:cascade}.

\subsection{MHD Inertial Range: From Driving Scales to the Collisional Transition}
\label{sec:coll}

The turbulence in astrophysical environments is typically driven by
some external mechanism, often a violent event or large-scale
instability, that generates plasma motions at some large scale, $L \gg
\rho_i$.  This energy injection scale, often denoted  the outer
scale of the turbulence, is an important characteristic of any
turbulent astrophysical system, and is conveniently parameterized by
the wavenumber, $k_0\sim 1/L$.  For the investigation of kinetic
turbulence, a convenient dimensionless measure of the driving scale is
the \emph{driving wavenumber}, $k_0 \rho_i$, where $k_0 \rho_i\ll 1$
indicates that the turbulence is driven at large scale compared to the
ion Larmor radius. It is generally assumed, in the absence of
arguments to the contrary, that the turbulence is driven isotropically
with respect to the magnetic field, so that the perpendicular and
parallel components of the driving wavevector are equal, $k_{\parallel
0} \sim k_{\perp 0}
\sim k_0 $.

If the MHD approximation is satisfied for the turbulent dynamics of
the plasma at the driving scale, then the large scale section of the
MHD inertial range is described by MHD turbulence theory
\citep{Sridhar:1994,Goldreich:1995,Galtier:2000,Lithwick:2001,Boldyrev:2006}.
If the amplitude of the driven turbulent velocities are comparable to
the \Alfven velocity in the magnetized plasma, then a cascade of
strong MHD turbulence arises to transfer energy nonlinearly to higher
wavenumbers; for smaller amplitudes, weak MHD turbulence will be
generated \citep{Sridhar:1994,Goldreich:1995}. Since most turbulent
astrophysical environments are believed to be driven strongly, and
weak turbulence eventually transitions to strong turbulence as the
cascade progresses \citep{Sridhar:1994}, we focus here on the case of
strong MHD turbulence.  In general, the finite-amplitude turbulent
fluctuations may be considered to be a mixture of the three
propagating MHD wave modes, the incompressible Alfv\'en waves and
compressible fast and slow waves, as well as the non-propagating
entropy mode. The nature of the turbulent cascades of these various
characteristic fluctuations have been elucidated by numerical
simulations of MHD turbulence: the fast waves cascade isotropically in
wavevector space, while the \Alfven waves, slow waves and entropy mode
fluctuations cascade anisotropically according to the condition of
critical balance \citep{Maron:2001,Cho:2003}.

The fast wave cascade produces an isotropic one-dimensional magnetic
energy spectrum $E_B(k) \propto k^{-3/2}$, as observed in simulations
\citep{Cho:2003}.  Two competing models exist that describe the nature 
of strong MHD turbulence for \Alfven waves, the Goldreich-Sridhar
model \citep{Goldreich:1995} and the Boldyrev model
\citep{Boldyrev:2006}.  The magnetic energy spectrum of the \Alfvenic
turbulent cascade is predicted to scale as $E_B \propto k_\perp ^p$,
where the spectral index is $p=-5/3$ in the Goldreich-Sridhar model
\citep{Goldreich:1995} and  $p=-3/2$ in the Boldyrev model
\citep{Boldyrev:2006}. The anisotropy of the \Alfvenic cascade, 
for isotropic driving at wavenumber $k_0$, is given by $k_\parallel
=k_0^{1-q} k_\perp^q$, where $q=2/3$ in the Goldreich-Sridhar model
\citep{Goldreich:1995}, and $q=1/2$ in the Boldyrev model
\citep{Boldyrev:2006}. The slow waves and entropy modes are passively 
cascaded by the \Alfven waves, and therefore adopt the same spectrum
and anisotropic distribution of power as the \Alfven waves
\citep{Maron:2001,Lithwick:2001}. For anisotropic turbulent fluctuations with
$k_\perp \gg k_\parallel$, the frequencies of the fast wave
fluctuations, which scale as $\omega \propto k$, are generally much
higher than the frequencies of the \Alfven and slow wave fluctuations,
which scale as $\omega \propto k_\parallel$, so the dynamics of the
fast wave cascade are expected to decouple from the dynamics of the
\Alfven and slow wave cascades  \citep{Lithwick:2001,Howes:2012a}.

The turbulent cascade transfers energy nonlinearly to higher
wavenumber fluctuations, as dictated by the MHD turbulence theory,
until the parallel wavenumber reaches the transition from collisional
to collisionless dynamics, $k_\parallel \lambda_i \sim 1$. The
perpendicular wavenumber, $k_{\perp c}$, that corresponds to
$k_\parallel \lambda_i \sim 1$, differs for the anisotropic \Alfven
wave cascade and the isotropic fast wave cascade. For the anisotropic
\Alfvenic cascade, the perpendicular wavenumber of this  transition is given
by $k_{\perp c} \sim k_0 (k_0 \lambda_i)^{-1/q}$, whereas, for the
isotropic fast wave cascade, it is given by $k_{\perp c} \sim k_0 (k_0
\lambda_i)^{-1}$, or more simply $k_{\perp c} \lambda_i\sim 1$. Since
$q<1$, this means that the fast wave cascade reaches the collisional
transition first, at a smaller wavenumber than the \Alfven wave
cascade. 

For many astrophysical plasmas, the transition for both fast and
\Alfven waves occurs within the MHD inertial range, $k_{\perp c }
\rho_i < 1$. The compressible fast waves, slow waves, and entropy modes 
undergo strong collisional damping by ion viscosity
\citep{Braginskii:1965} in the moderately collisional conditions at
$k_\perp \sim k_{\perp c}$.  Any energy in the compressible turbulent
fluctuations that passes through this transition is expected to be
transferred nonlinearly to the kinetic counterparts of the MHD fast and
slow waves \citep{Klein:2012} in the weakly collisional conditions at
wavenumbers $k_{\perp} \gg k_{\perp c }$ \citep{Schekochihin:2009}.
The \Alfven waves are incompressible, involving no motions parallel to
the magnetic field, so they are essentially unaffected by the
transition in collisionality, and the \Alfven wave cascade continues
unabated to higher wavenumbers, $k_{\perp} \gg k_{\perp c }$.

\subsection{MHD Inertial Range: From the Collisional Transition to the Ion Larmor Radius}

Critical balance predicts a scale-dependent wavevector anisotropy
given by $k_\perp/k_\parallel = (k_\perp /k_0 )^{1-q}$, where $q<1$
for either the Goldreich-Sridhar or Boldyrev models. Therefore, at
perpendicular wavenumbers within the MHD inertial range sufficiently
higher than the driving wavenumber, $k_\perp \gg k_0$, the \Alfvenic
fluctuations become anisotropic in the sense that $k_\perp \gg
k_\parallel$. In the limit of the MHD inertial range $k_\perp \rho_i
\ll 1$, the kinetic dynamics of these anisotropic \Alfvenic
fluctuations is described rigorously by the equations of reduced MHD
\citep{Strauss:1976}, and the \Alfven wave cascade remains undamped 
down to the perpendicular scale of the ion Larmor radius, $k_\perp \rho_i \sim 1$
\citep{Schekochihin:2009}. It has also been shown that the slow wave 
and \Alfven wave cascades do not exchange energy in the MHD inertial
range \citep{Schekochihin:2009}, and the fast waves likewise are not
expected to exchange energy with the \Alfven waves due to the mismatch
in frequency, as discussed in \secref{sec:coll}.  Therefore, the
dynamics of the \Alfvenic cascade throughout the MHD inertial range is
correctly described by the MHD turbulence theory, even at the weakly
collisional scales, $k_\perp \gg k_{\perp c}$. 

The magnetic energy spectrum in the solar wind seems to bear this
out. Spacecraft measurements in the super-\Alfvenic solar wind are
generally interpreted by assuming the Taylor hypothesis
\citep{Taylor:1938}, that frequency of measured temporal fluctuations
is directly related to the wavenumber of spatial variations that are
swept past the spacecraft.  At the frequencies $f \lesssim 0.4$~Hz,
corresponding to spatial scales larger than the ion Larmor radius, the
magnetic energy spectrum in the solar wind has a spectral index of
approximately $-5/3$ \citep{Goldstein:1995}, apparently consistent
with the prediction of the Goldreich-Sridhar theory for strong MHD
turbulence. It is worth noting, however, that the velocity spectrum
was found have a spectral index closer to $-3/2$
\citep{Podesta:2007}, in conflict with the Goldreich-Sridhar model. 
Recent work on the evolution of the residual energy, $E_r=E_k-E_B$, in
MHD turbulence, however, suggests that these spectral indices may
indeed be consistent with the Boldyrev theory, and that the difference
in the spectral indices of the kinetic and magnetic energy spectra is
an inherent property of the MHD turbulent cascade
\citep{Boldyrev:2011,Boldyrev:2012a}.

The cascade of compressible turbulent fluctuations that passes through
the collisional transition will suffer moderate to strong
collisionless damping by the Landau resonance with the ions
\citep{Barnes:1966} at all higher wavenumbers, $k_\perp \gg k_{\perp
c}$. The damping of the compressible fluctuations in the moderate to
weakly collisional regimes at $k_\perp \gtrsim k_{\perp c}$ leads to
the theoretical prediction that compressible fluctuations will play a
subdominant role relative to the undamped \Alfvenic fluctuations in
turbulent astrophysical plasmas.  Studies of interstellar
scintillation \citep{Armstrong:1981,Armstrong:1995} show evidence for a power-law
spectrum of density fluctuations over 12 orders of magnitude in the
interstellar medium, suggesting that compressible fluctuations are not
entirely damped. But it is not possible from remote astrophysical
observations to deduce the relative contributions of compressible and
incompressible components of the turbulence. 

\emph{In situ} spacecraft measurements of turbulent fluctuations in
the solar wind, however, allow a direct determination. The entire
turbulent cascade in the solar wind, including the driving scales, is
weakly collisional, $\lambda_i/L \gg 1$, so spacecraft measurements
constrain role of compressible fluctuations in collisionless
conditions, $k_\perp \gg k_{\perp c}$.  Measurements show that the
turbulent fluctuations in the MHD inertial range appear to be
dominantly incompressible \citep{Tu:1995,Bruno:2005}, where the
incompressible motions have been shown to be Alfv\'enic in nature
\citep{Belcher:1971}.  The compressible fluctuations generally
contribute less than 10\% of the turbulent magnetic energy
\citep{Tu:1995,Bruno:2005}.  These compressible fluctuations have
typically been interpreted as a possible mixture of fast MHD waves and
pressure balanced structures (PBSs)
\citep{Tu:1995,Bruno:2005}, where the latter are equivalent to
non-propagating slow mode fluctuations with $k_\parallel =0$
\citep{Tu:1994,Kellogg:2005}.
Note, however, that a recent study using a novel method of synthetic
spacecraft data \citep{Klein:2012} suggests that these compressible
fluctuations are not associated with the kinetic counterpart of the
fast MHD wave, but rather consist of an anisotropic distribution of
kinetic slow wave fluctuations \citep{Howes:2012a}.  Clearly, more
investigation of the kinetic physics of compressible turbulent
fluctuations in astrophysical environments, including their damping
via collisional and collisionless mechanisms and the resulting plasma
heating, is needed. Nonetheless, since only a small fraction of the
turbulent energy appears to be associated with the compressible
fluctuations, we focus our attention henceforth on the dominant
\Alfvenic turbulent fluctuations, as they reach the  perpendicular
scale of the ion Larmor radius, $k_\perp \rho_i \sim 1$.

\subsection{Transition at the Ion Larmor Radius}
\label{sec:transition}

Spacecraft measurements of turbulence in the solar wind demonstrate
that the $-5/3$ scaling of the magnetic energy spectrum in the MHD
inertial range breaks at a frequency around $f \sim 0.4$~Hz, leading
to a steeper spectrum at higher frequencies in the kinetic dissipation
range. Numerous observational studies have attempted to correlate the
position of the break with a characteristic plasma time or length
scale, such as the ion cyclotron frequency, the ion Larmor radius, or
the ion inertial length
\citep{Goldstein:1994,Leamon:1998a,Leamon:1999,Leamon:2000,Smith:2001b,Perri:2010b,Smith:2012,Bourouaine:2012},
but contradictory results have been found. Establishing a convincing
correlation has likely been elusive because three competing effects
may contribute to the dynamics at this transition between the MHD
inertial range and kinetic dissipation range: (1) the transition from
non-dispersive to dispersive linear wave physics as the ions decouple
from the turbulent electromagnetic fluctuations; (2) a peak in the ion
kinetic damping; and (3) the possible role of kinetic instabilities,
such as temperature anisotropy instabilities \citep{Bale:2009}, in
generating electromagnetic fluctuations at this scale.

Based on theoretical considerations of the kinetic plasma physics, the
kinetic turbulence model presented here predicts that the transition
between the relatively well understood MHD inertial range and the
significantly more controversial kinetic dissipation range occurs at
the perpendicular scale of the ion Larmor radius, $k_\perp \rho_i \sim
1$ \citep{Howes:2008c,Howes:2008b,Schekochihin:2009,Howes:2011b}.  The
boundary conditions (in wavevector space) for the nonlinear transfer
of energy into the kinetic dissipation range are given by the nature
of the turbulent fluctuations at the end of the MHD inertial range. At
this transition at $k_\perp \rho_i \sim 1$, the wavevector anisotropy
of \Alfvenic turbulent fluctuations is given by $k_\perp/k_\parallel
\sim (k_0 \rho_i)^{q-1}$, so for a sufficiently large MHD inertial
range, $k_0\rho_i \ll 1$, this implies $k_\perp \gg k_\parallel$ since
$q<1$ \citep{Goldreich:1995,Boldyrev:2006}. This significant
wavevector anisotropy at the transition is supported by
multi-spacecraft measurements of turbulence in the near-earth Solar
wind \citep{Sahraoui:2010b}. It follows that, beyond this transition,
the characteristic wavevector of the fluctuations satisfies $k_\perp
\rho_i \gtrsim 1$ and $k_\parallel \rho_i \ll 1$; the
\Alfvenic solution of linear kinetic theory with such a wavevector
is the kinetic \Alfven wave \citep{Hasegawa:1989,Stix:1992}.
Therefore, the \Alfven waves of the MHD inertial range are predicted
to transfer their energy, via nonlinear interactions at the transition
$k_\perp \rho_i \sim 1$, to kinetic
\Alfven waves \citep{Leamon:1998a,Gruzinov:1998,Leamon:1999,Quataert:1999,
Howes:2008b,Schekochihin:2009}.  Nonlinear gyrokinetic simulations of
this transition appear to support this hypothesis \citep{Howes:2008a},
reproducing the qualitative changes in the electric and magnetic field
energy spectra measured in the solar wind at the scale of the spectral
break \citep{Bale:2005}.

Another important effect that occurs at the transition at $k_\perp
\rho_i \sim 1$ is a peak in the collisionless damping rate of the
electromagnetic fluctuations due to the Landau resonance with the ions
\citep{Leamon:1998a,Leamon:1999,Leamon:2000,Howes:2008b,Howes:2008c,
Schekochihin:2009,Howes:2011b}.  This ion kinetic damping becomes
increasingly strong as the ion plasma beta increases, and is generally
non-negligible for plasmas with beta of order unity or larger,
$\beta_i \gtrsim 1$, leading to a significant fraction of the
dissipated turbulent energy heating the ions
\citep{Howes:2010d,Howes:2011c}, in approximate agreement 
with empirical estimates of the plasma heating in the solar wind
\citep{Cranmer:2009,Breech:2009}.  Some measurements of the magnetic
energy spectrum in the dissipation range of the solar wind show a
significant steepening to a slope of approximately $-4$ at the ion
scales, flattening to $-2.8$ spectrum further into the dissipation
range \citep{Sahraoui:2010b}, evidence suggesting significant ion kinetic
damping.

In a steady-state kinetic turbulent cascade, the turbulent energy
reaching the transition at $k_\perp \rho_i \sim 1$ that is not damped
at that scale will carry on, launching a turbulent cascade of kinetic
\Alfven waves in the kinetic dissipation range at $k_\perp \rho_i
\gtrsim 1$.  Although the \Alfven and slow wave cascades do not
exchange energy in the MHD inertial range, they may exchange energy at
this transition \citep{Schekochihin:2009}, so it is possible that the
kinetic \Alfven wave cascade can gain energy that is transferred
nonlinearly from compressible fluctuations in the MHD inertial range.

\subsection{Kinetic Dissipation Range: Between the Ion and Electron Larmor Radius}

Although direct spacecraft measurements of the kinetic dissipation
range of turbulence in the near-Earth solar wind have been possible
for more than a decade, the nature of the turbulent fluctuations in
this regime remains a controversial topic. Characterizing these
fluctuations is one of the key goals in heliospheric physics today,
especially because the relevant physical dissipation mechanisms that
ultimately lead to heating of the plasma depend strongly on the nature
of the turbulent fluctuations themselves.

Many early investigations of the dissipation range in solar wind
turbulence implicitly assumed that the turbulent fluctuations in the
dissipation range are related to the linear wave modes in the plasma.
Two main hypotheses have been proposed, that the turbulence is
composed of either kinetic \Alfven waves
\citep{Leamon:1998a,Gruzinov:1998,Leamon:1999,Quataert:1999,
Howes:2008b,Schekochihin:2009} or whistler waves
\citep{Stawicki:2001,Gary:2010,Narita:2010b}. Although these two
possibilities generally remain the leading candidates, several other
possibilities have been suggested: ion Bernstein waves
\citep{Sahraoui:2012}, ion cyclotron waves \citep{Jian:2009},
non-propagating pressure balanced structures (PBSs), or inherently
nonlinear structures, particularly highly intermittent coherent
structures and current sheets \citep{Servidio:2011b}.

Direct spacecraft measurements of turbulence in the solar wind at the
frequencies $f \gtrsim 1$~Hz, corresponding to the kinetic dissipation
range, provide important constraints on the nature of the turbulent
fluctuations.  A number of recent studies employing high temporal
resolution spacecraft measurements have found a nearly
power-law scaling of the magnetic energy spectrum between the ion and
electron scales with a spectral index of approximately $-2.8$
\citep{Sahraoui:2009,Kiyani:2009,Alexandrova:2009,Chen:2010b,Sahraoui:2010b}.
These observations of the turbulence over the dissipation range scales
raise two important questions that any model for kinetic turbulence
must answer: (1) What causes the magnetic energy spectrum to steepen
in the dissipation range; and (2) Does significant dissipation of the
turbulent fluctuations occur between the ion and electron scales?

In the model of the kinetic turbulent cascade, the boundary conditions
in wavevector space determined by the anisotropic \Alfvenic cascade
through the MHD inertial range suggest that the turbulent energy is
transferred nonlinearly to a cascade of kinetic \Alfven waves in the
kinetic dissipation range, as discussed in \secref{sec:transition}.
Here we describe the properties of the kinetic
\Alfven wave cascade at perpendicular scales below the ion Larmor
radius, $k_\perp \rho_i \gtrsim 1$. 

Although MHD \Alfven waves are non-dispersive, kinetic \Alfven waves
become dispersive due to the averaging of the ion response over the
finite ion Larmor radius, a physical effect that increasingly
decouples the ions from the electromagnetic fluctuations with
wavevectors satisfying $k_\perp \rho_i \gtrsim 1$
\citep{Hollweg:1999,Schekochihin:2009}.  A useful formula combining
the linear frequency in the \Alfven and kinetic \Alfven wave regimes
\cite{Howes:2006,Schekochihin:2009} is given by
\begin{equation}
\omega = k_\parallel v_A \sqrt{1 + \frac{(k_\perp \rho_i)^2}{\beta_i + 2/(1+T_e/T_i)}}
\label{eq:kaw}
\end{equation}
In addition, the kinetic \Alfven wave is significantly compressible,
generating a non-zero parallel magnetic field fluctuation, $\delta
B_\parallel$, particularly in the limit of low to moderate plasma
beta, $\beta_i \lesssim1$ \citep{Hollweg:1999,TenBarge:2012b}.

The model for kinetic turbulence predicts the quantitative scaling of the
magnetic energy spectrum and the wavevector anisotropy for the kinetic
\Alfven wave cascade. The Kolmogorov hypothesis---that  the energy 
transfer rate is constant due to local (in wavenumber space) nonlinear
interactions---can be used to predict the magnetic energy spectrum for
the kinetic \Alfven wave cascade in the absence of significant
dissipation. For $k_\perp \rho_i \gg 1$, the linear wave frequency
increases due to dispersion, yielding a scaling $\omega \propto
k_\parallel k_\perp $. This leads to more rapid nonlinear energy
transfer, steepening the magnetic energy spectrum to a predicted
scaling $E_B \propto k_\perp^{-7/3}$ when dissipation is neglected
\citep{Biskamp:1999,Cho:2004,Krishan:2004,Shaikh:2005,
Galtier:2006,Howes:2008b,Schekochihin:2009}. Extending the concept of
critical balance---that the linear wave frequency and nonlinear energy
transfer frequency remain in balance---to the kinetic \Alfven wave
regime leads to a predicted wavevector anisotropy given by $
k_\parallel \propto k_\perp^{1/3}$
\citep{Cho:2004,Howes:2008b,Schekochihin:2009}.

In addition to the effects of wave dispersion, collisionless damping
via wave-particle interactions can also play an important role in
kinetic turbulence for all scales $k_\perp \rho_i \gtrsim 1$. In
addition to the peak in ion Landau damping at $k_\perp \rho_i \sim 1$
discussed in \secref{sec:transition}, electron Landau damping may also
play a significant role for all scales $k_\perp \rho_i \gtrsim 1$,
becoming increasingly strong as the perpendicular wavenumber increases
\citep{Howes:2008b}. Although early models of the turbulent energy
cascade in the kinetic dissipation range suggested that such strong
collisionless Landau damping would lead to an exponential cutoff of
the spectrum before reaching the perpendicular scale of the electron
Larmor radius, $k_\perp \rho_e \sim 1$
\citep{Howes:2008b,Podesta:2010a}, subsequent 
solar wind observations called this prediction into question
\citep{Sahraoui:2009,Kiyani:2009,Alexandrova:2009,Chen:2010b,Sahraoui:2010b} 
and recent kinetic numerical simulations have demonstrated that this
idea is incorrect \citep{Howes:2011a}.

In addition to collisionless damping via the Landau resonance, if the
kinetic \Alfven wave frequency reaches the ion cyclotron frequency,
$\omega \rightarrow \Omega_i$, collisionless damping may occur via the
cyclotron resonance with the ions. However, the very large MHD
inertial range typical of astrophysical plasma turbulence leads to
highly anisotropic fluctuations at small scales, $k_\parallel \ll
k_\perp$, so the kinetic \Alfven wave frequency typically remains very
small compared to the ion cyclotron frequency, $\omega \ll \Omega_i$.
Therefore, ion cyclotron damping is not predicted to play a strong
role in the dissipation of astrophysical turbulence
\citep{Howes:2008b}, with a few exceptions, such as the inner 
heliosphere \citep{Howes:2011c}. 

There is significant evidence accumulating in support of a kinetic
\Alfven wave cascade at the perpendicular scales between the electron
and ion Larmor radius, but there also remains observational evidence
that appears to be unexplained by this model.  The scaling predictions
for the kinetic \Alfven wave cascade in the absence of dissipation
have been corroborated by simulations using electron MHD, a fluid
limit which describes the dynamics of kinetic
\Alfven waves in the limit $k_\parallel \ll k_\perp$, but does not
resolve the physics of collisionless dissipation.  Specifically, these
simulations reproduce the predicted magnetic energy scaling, $E_B
\propto k_\perp^{-7/3}$
\citep{Biskamp:1999,Cho:2004,Cho:2009,Shaikh:2009}, and wavevector anisotropy, 
$ k_\parallel \propto k_\perp^{1/3}$ \citep{Cho:2004,Cho:2009}. The
magnetic energy spectrum from these fluid simulations, however, is not
consistent with the observed spectral index of approximately $-2.8$
\citep{Sahraoui:2009,Kiyani:2009,Alexandrova:2009,Chen:2010b,Sahraoui:2010b}.

It has been recently suggested that the combined effects of
collisionless dissipation and nonlocal energy transfer can lead to a
further steepening of the magnetic energy spectrum beyond
$k_\perp^{-7/3}$ for scales $k_\perp
\rho_i \gtrsim 1$ \citep{Howes:2011b}.  For a hydrogenic plasma of protons 
and electrons, the dynamic range between the ion and electron Larmor
radius for unity temperature ratio is $\rho_i/\rho_e \simeq 43$, so there
is little room for an asymptotic range of perpendicular scales
satisfying the requirements $1/\rho_i \ll k_\perp \ll
1/\rho_e$. Therefore, it should come as no surprise that a
self-similar spectrum with a spectral index of $-7/3$ is not
observed---throughout the range of perpendicular scales between the
ion and electron Larmor radius, the transition at $k_\perp \rho_i \sim
1$ and strong kinetic dissipation at $k_\perp
\rho_e \sim 1$ may significantly affect the turbulent dynamics.
 
Nonlinear gyrokinetic simulations of turbulence in the kinetic
dissipation range seem to bear this out. A simulation over the entire
range of scales from the ion to the electron Larmor radius, which
resolves the physics of collisionless ion and electron damping,
produces a nearly power-law magnetic energy spectrum with a spectral
index of $-2.8$, in remarkable quantitative agreement with the solar
wind measurements \citep{Howes:2011a}.  Additional gyrokinetic
simulations support the predicted scaling of the wavevector
anisotropy, $k_\parallel \propto k_\perp^{1/3}$
\citep{TenBarge:2012a,TenBarge:2013b}.

Direct spacecraft measurements of dissipation range turbulence in the
solar wind have yielded other lines of evidence in support of or in
conflict with the model of a kinetic \Alfven wave cascade.  A
$k$-filtering analysis of multi-spacecraft data from the Cluster
mission establishes that the plasma-frame fluctuation frequencies are
consistent with linear dispersion relation of the kinetic \Alfven wave
and inconsistent with that of the whistler wave
\citep{Sahraoui:2010b}. A study combining measurements of the ratio of
electric to magnetic field fluctuation amplitudes and of the magnetic
compressibility have shown that the small-scale fluctuations agree
well with predictions for kinetic \Alfven waves and are inconsistent
with that for whistler waves \citep{Salem:2012}. An examination the
compressibility of turbulent fluctuations in the weakly collisional
plasma in the MHD inertial range finds evidence of negligible energy
in the fast wave mode, suggesting that all large-scale turbulent energy is
transferred, via the anisotropic \Alfvenic cascade, to kinetic \Alfven
waves, with little energy coupling to whistler waves
\citep{Howes:2012a,Klein:2012}.  Investigations of the magnetic
helicity of turbulent fluctuations as a function of the angle of the
wavevector with respect to the local magnetic field direction finds a
broad region of positive helicity at oblique angles
\citep{He:2011,Podesta:2011a}, as expected for kinetic \Alfven waves
\citep{Howes:2010a}, but a small region corresponding to  nearly parallel 
wavevectors that is consistent with either ion cyclotron waves or
whistler waves \citep{He:2011,Podesta:2011a}.

The presence of either ion cyclotron or whistler waves with nearly
parallel wavevectors is not explained by the model for kinetic
\Alfven wave turbulence, but these fluctuations may be driven by the action 
of kinetic temperature anisotropy instabilities in the spherically
expanding solar wind \citep{Bale:2009}. These instabilities typically
generate relatively isotropic fluctuations (with respect to the local
mean magnetic field direction), having wavevector components $k_\perp \rho_i 
\sim k_\parallel \rho_i \sim 1$.  Since the anisotropic \Alfvenic cascade
produces fluctuations with $k_\parallel \ll k_\perp$, and since
\Alfvenic frequencies in \eqref{eq:kaw} scale linearly with the parallel component,
$\omega \propto k_\parallel$, these anisotropy-driven fluctuations are
expected to have a much higher frequency and to occupy a different
regime of wavevector space than turbulent fluctuations of the
turbulent cascade.  Therefore, it is reasonable to expect that any
kinetic-instability-driven fluctuations may persist without
significantly affecting, or being significantly affected by, the
turbulent fluctuations of the anisotropic \Alfvenic cascade.

In conclusion, although the nature of the kinetic turbulence at the
perpendicular scales between the electron and ion Larmor radius has
not been established conclusively, there appears to be significant
evidence for an anisotropic cascade of kinetic \Alfven waves in the
kinetic dissipation range. Collisionless dissipation via the Landau
resonance with the electrons appears to be play a non-negligible role
in steepening the magnetic energy spectrum beyond the dissipationless
prediction.  But this damping is not strong enough to halt the
cascade, so the kinetic turbulence continues down to the perpendicular
scale of the electron Larmor radius, at which point strong kinetic
dissipation can effectively terminate the turbulent cascade.

\subsection{Kinetic Dissipation Range: Termination at Electron Larmor Radius}
Ultimately, the kinetic turbulent cascade reaches the perpendicular
scale of the electron Larmor radius, $k_\perp \rho_e \sim 1$.  At this
scale, the linear collisionless damping rate due to electron Landau
damping reaches a value $\gamma/\omega \sim 1$, sufficiently strong
that the turbulent magnetic energy cascade is terminated. A simplified
analytical treatment of the turbulent cascade undergoing this
dissipation suggests the spectrum will develop an exponential fall-off
$E_B \propto k_\perp^{-2.8} \exp(-k_\perp \rho_e)$ setting in at the
perpendicular scale of the electron Larmor radius, $k_\perp \rho_e
\sim 1$ \citep{Terry:2012}. As the amplitudes of the turbulent 
fluctuations are diminished by damping, the strong kinetic \Alfven
wave turbulence eventually drops below critical balance and becomes
weak dissipating kinetic \Alfven wave turbulence
\citep{Howes:2011b}. It has been conjectured that the 
transition back to weak turbulence leads to an inhibition of the
parallel cascade, so the parallel number of the fluctuations remains
constant \citep{Howes:2011b}, as shown in in \figref{fig:kinetic}(b).

A recent study of a sample of 100 solar wind magnetic energy spectra
at the electron scales shows that all of these spectra may be fit by
an empirical form $E_B \propto k_\perp^{\alpha}
\exp(-k_\perp \rho_e)$, where $-2.5 \ge \alpha \ge -2.8$ \citep{Alexandrova:2012}.
A nonlinear gyrokinetic simulation of the turbulence over the range
$0.12 \le k_\perp \rho_e \le 2.5$ yields an energy spectrum
demonstrating an exponential fall-off that is quantitatively fit by
$E_B \propto k_\perp^{-2.8} \exp(-k_\perp \rho_e)$, further supporting
the model of the kinetic turbulent cascade \citep{TenBarge:2013b}.  A
refined model of the turbulent cascade
\citep{Howes:2011b}---incorporating the weakening of the nonlinear turbulent
energy transfer due to dissipation, the effect of nonlocal energy
transfer, and the linear collisionless damping via the Landau
resonance---fits the shape of the spectrum well.  This provides
compelling evidence that collisionless damping is the dominant
mechanism for the dissipation of the kinetic turbulent cascade,
marking the end of the kinetic dissipation range
\citep{TenBarge:2013b}.  

A number of recent works have suggested instead that dissipation in
current sheets is the dominant dissipation mechanism for plasma
turbulence, based on fluid simulations using MHD
\citep{Dmitruk:2004,Servidio:2009,Servidio:2010,Servidio:2011b}
and Hall MHD \citep{Dmitruk:2006b}, hybrid simulations with kinetic
ions and fluid electrons
\citep{Parashar:2009,Parashar:2010,Markovskii:2011,Servidio:2012}, and
observational studies of large-scale discontinuities in the solar wind
\citep{Osman:2011,Osman:2012a}. However, all the numerical work upon
which this conclusion has been based employ a fluid description of the
electrons which does not resolve the dominant collisionless
dissipation mechanism of Landau damping. In a collisionless plasma,
current sheets supporting small-scale reconnection with a guiding
magnetic field are expected to form at the perpendicular scale of the
electron Larmor radius, $k_\perp\rho_e \sim 1$ \citep{Birn:2006}.  A
recent gyrokinetic simulation over the range of electron scales $0.12
\le k_\perp \rho_e \le 2.5$, which resolves both  collisionless
electron Landau damping and the formation of current sheets at
$k_\perp\rho_e \sim 1$, finds dissipation via current sheets to be
sub-dominant compared to linear collisionless damping
\citep{TenBarge:2013a}.

These results establish fairly secure observational and numerical
grounds that the electromagnetic fluctuations of the kinetic turbulent
cascade are dissipated at the perpendicular scale of the electron
Larmor radius, $k_\perp\rho_e \sim 1$. Further work is ongoing to
identify the dominant physical mechanisms for the dissipation of these
turbulent electromagnetic fluctuations.  Although this dissipation
terminates the kinetic dissipation range of electromagnetic
fluctuations at electron scales, there remains the final matter of
identifying the physical mechanism mediating the conversion of this
turbulent fluctuation energy irreversibly to thermodynamic heat in a
weakly collisional astrophysical plasma.

\subsection{Irreversible Heating Via the Ion and Electron Entropy Cascades}

At the perpendicular scales of the ion and electron Larmor radius,
collisionless wave-particle interactions via the Landau resonance damp
the turbulent electromagnetic fluctuations. In the absence of
collisions, this process conserves a generalized energy, generating
nonthermal structure in velocity space of the corresponding plasma
particle distribution functions
\citep{Howes:2008c,Schekochihin:2009,Plunk:2010}. Boltzmann's $H$-theorem 
dictates that, in a kinetic plasma, collisions are required to increase
the entropy and therefore achieve irreversible thermodynamic heating
\citep{Howes:2006}.  In the weakly collisional plasmas of 
astrophysical environments, an \emph{entropy cascade}---a
nonlinear phase mixing process \citep{Dorland:1993} that drives a dual
cascade in physical and velocity space---mediates the transfer, at
sub-Larmor radius scales, of the nonthermal free energy in the
particle distribution functions to sufficiently small scales in
velocity space that arbitrarily weak collisions can manifest
irreversibility, increasing the entropy and thermodynamically heating
the plasma \citep{Schekochihin:2009,Plunk:2010}. This inherently
kinetic physical mechanism represents the final element of the kinetic
turbulent cascade, governing the final transition of the turbulent
energy to its ultimate fate as plasma heat.

The ion entropy cascade in two-dimensional plasma systems (in the
plane perpendicular to the local mean magnetic field) has been
thoroughly examined theoretically \citep{Schekochihin:2009,Plunk:2010}
and verified in gyrokinetic numerical simulations
\citep{Tatsuno:2009,Plunk:2011}. In the inherently three-dimensional 
system of \Alfvenic plasma turbulence \citep{Howes:2011a}, the effects
of the ion entropy cascade in generating structure in the
perpendicular component of velocity space \citep{Howes:2008c} and in
manifesting ion heating at physical scales well below the peak in the
collisionless ion damping \citep{Howes:2011a} have been identified
numerically. Yet, a thorough analysis of the ion and electron entropy
cascades in kinetic turbulence remains to be undertaken.


\section{Conclusion}
Turbulence is found ubiquitously throughout the universe, playing a
governing role in the conversion of the energy of large-scale motions
to astrophysical plasma heat. Extending our understanding of
astrophysical turbulence from the limited theory of MHD turbulence to
the more comprehensive theory of kinetic turbulence opens up the
possibility of ultimately achieving a predictive capability to
determine the plasma heating due to the dissipation of
turbulence. This chapter has outlined a theoretical model of the
kinetic turbulent cascade describing the flow of energy from the large
driving scales, through the MHD inertial range, to the transition at
the ion Larmor radius, and into the kinetic dissipation range, where
the energy is ultimately converted to plasma heat. Although
significant progress has already been made, much research remains to
be done to refine the kinetic turbulence model and test its
predictions using numerical simulations and observational data.

Since kinetic turbulence includes a number of the physical processes
that are inherently kinetic---such as collisionless wave-particle
interactions and the entropy cascade---kinetic numerical simulations
will play an essential role in testing the predictions of this model.
The higher dimensionality of kinetic systems---in general, requiring
three dimensions in physical space and three dimensions in velocity
space---demands a huge investment of computational resources to
perform numerical simulations. It is tempting to reduce the
dimensionality in physical space to two-dimensions to lower the
computational costs, but doing so fundamentally limits the
applicability of the results to turbulent astrophysical plasmas.
The reason is because the anisotropic \Alfvenic turbulence dominating
the kinetic turbulent cascade is inherently three-dimensional in
physical space: the dominant nonlinearity responsible for the
turbulent cascade requires both dimensions perpendicular to the
magnetic field, and \Alfvenic fluctuations require variation in the
parallel dimension \citep{Howes:2011a}. Therefore, kinetic simulation
results can only be directly compared to astrophysical systems if
physical space is modeled in three dimensions.

Observational tests of the kinetic turbulence model should exploit
intuition from kinetic plasma theory to unravel the dependence of the
turbulent properties on the plasma parameters. The suitably normalized
MHD linear dispersion relation depends only on two dimensionless
parameters, $\hat{\omega}_{\mbox{MHD}} = \omega/(k v_A) =
\hat{\omega}_{\mbox{MHD}}(\beta,\theta)$: the plasma beta $\beta$, and the 
angle between the wavevector and the magnetic field $\theta$
\citep{Klein:2012}. In contrast, the linear physics of Vlasov-Maxwell
kinetic theory depends on five dimensionless parameters,
$\tilde{\omega}_{\mbox{VM}} =
\omega/\Omega_i = \tilde{\omega}_{\mbox{VM}}(k_\parallel \rho_i, k_\perp \rho_i, \beta_i ,T_i/T_e, v_{ti}/c)$: the normalized parallel wavenumber $k_\parallel \rho_i$, 
the normalized perpendicular wavenumber $k_\perp \rho_i$, the ion
plasma beta $\beta_i$, the ion-to-electron temperature ratio
$T_i/T_e$, and the ratio of the ion thermal velocity to the speed of
light $v_{ti}/c$ \citep{Stix:1992,Howes:2006}.\footnote{In the limit that the
turbulent astrophysical fluctuations satisfy the gyrokinetic
approximation, $k_\parallel \ll k_\perp$ and $\omega \ll \Omega_i$
\citep{Frieman:1982,Howes:2006,Schekochihin:2009}, the linear physics depends on only three 
dimensionless parameters, $\overline{\omega}_{\mbox{GK}} =
\omega/(k_\parallel v_A) = \overline{\omega}_{\mbox{GK}}(k_\perp
\rho_i, \beta_i ,T_i/T_e)$ \citep{Howes:2006}.}  The dynamical behavior of the 
kinetic plasma---for example, the frequencies, collisionless damping
rates, and eigenfunctions of the fluctuations---varies as these
dimensionless parameters are changed.  Therefore, since the kinetic
plasma physics depends on these parameters, observational
investigations of kinetic turbulence should strive to analyze
measurements in terms of the ion plasma beta $\beta_i$, the
ion-to-electron temperature ratio $T_i/T_e$, and length scales
normalized to a characteristic plasma  kinetic length scale, such as the
ion Larmor radius or ion inertial length.

\begin{figure}[top] 
\vspace*{2mm}
\begin{center}
\resizebox{10.cm}{!}{
\includegraphics*{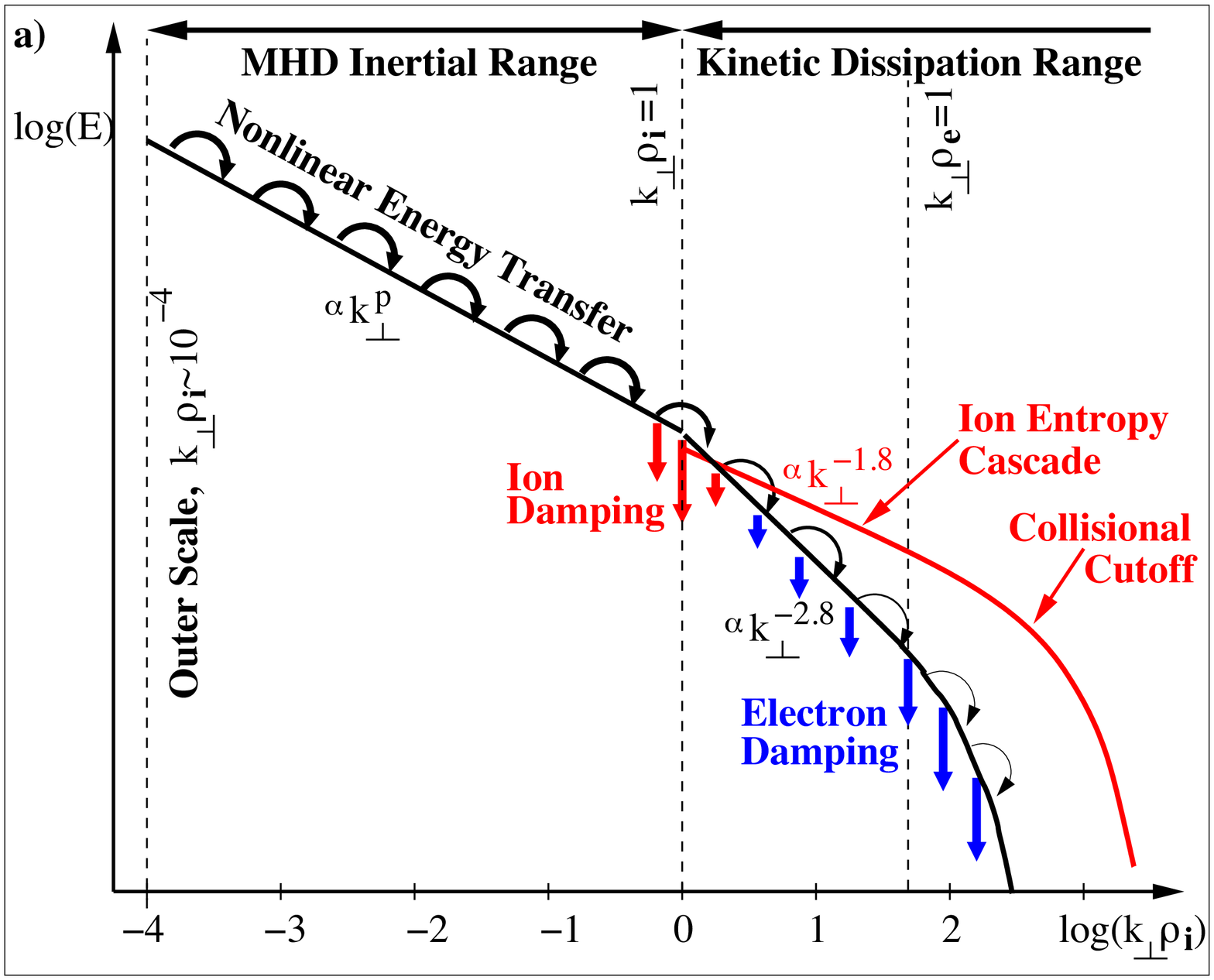}} 
\resizebox{10.cm}{!}{
\includegraphics*{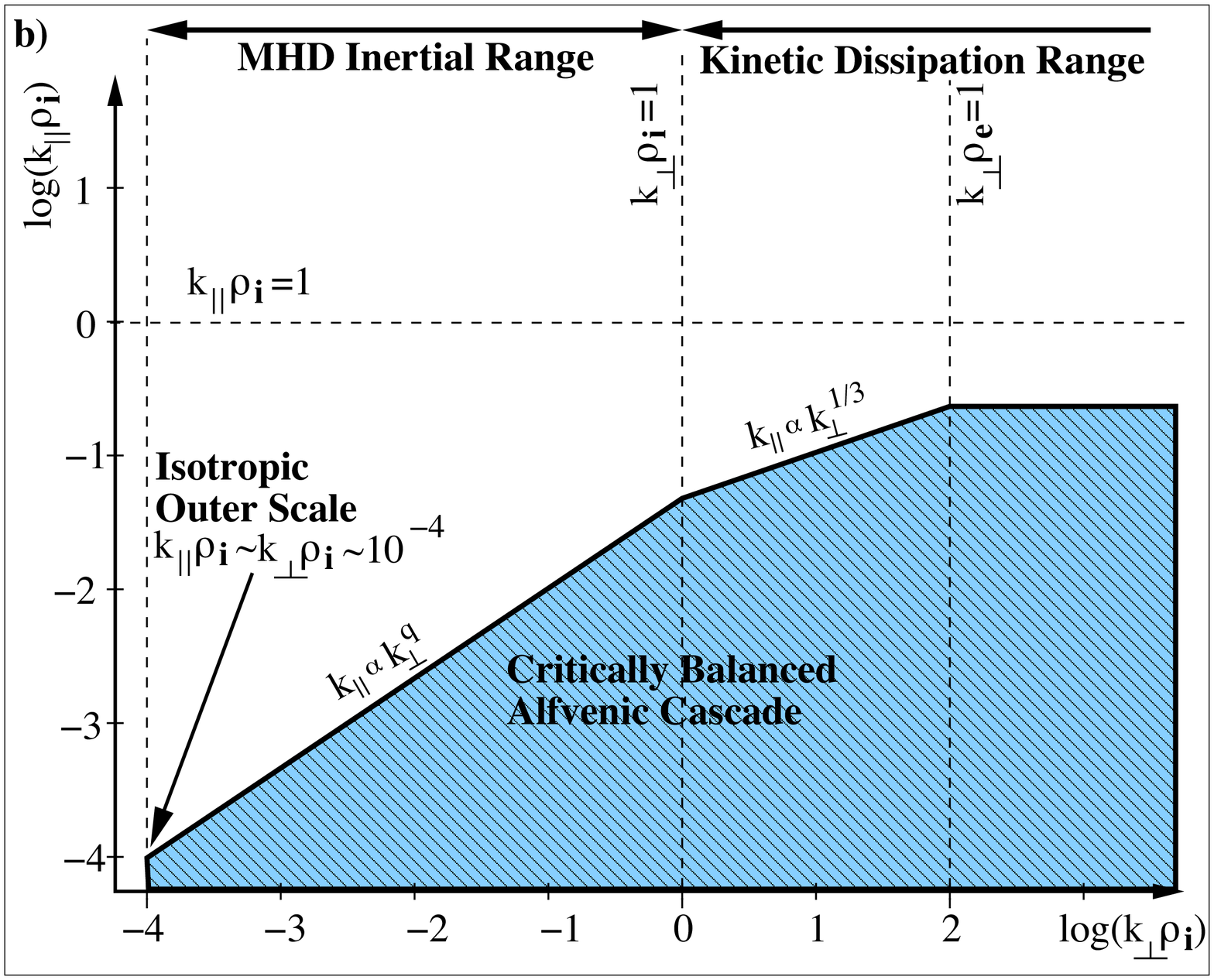}}
\end{center}
\caption{(a)~Diagram of the magnetic energy spectrum and 
ion entropy cascade in kinetic turbulence.  (b)~Anisotropic
distribution of power in $(k_\perp,k_\parallel)$ wavevector space in
kinetic turbulence. }
\label{fig:cascade} 
\end{figure}

Finally, we conclude this chapter on kinetic turbulence with a
schematic diagram that depicts the salient features of the kinetic
turbulent cascade, as shown in \figref{fig:cascade} for the case of
turbulence in the near-Earth solar wind (the relevant panel is noted
in parentheses in the description below). The turbulence is driven
isotropically at a large scale, corresponding to a normalized
wavenumber $k_0 \rho_i \sim 10^{-4}$ (a,b). Nonlinear interactions
serve to transfer the energy from this low driving wavenumber to
higher wavenumber through the MHD inertial range, generating a
magnetic energy spectrum scaling as $E_B \propto k_\perp^{p}$, where
$p=-5/3$ or $-3/2$ (a). This cascade is anisotropic in wavevector
space, such that turbulent fluctuations fill the shaded region below
$k_\parallel \propto k_\perp ^q$, where $q=2/3$ or $1/2$ (b). The
\Alfvenic turbulence transitions from the MHD inertial range to the
kinetic dissipation range at a perpendicular wavenumber $ k_\perp
\rho_i \sim 1$, a scale at which collisionless ion Landau damping
peaks (a). The energy transferred via wave-particle interactions to
the ion distribution function feeds the ion entropy cascade at
wavenumbers $ k_\perp \rho_i \gtrsim 1$, a dual cascade in physical
and velocity space that mediates the transfer of nonthermal structure
to sufficiently small scales in velocity space that weak collisions
can thermalize the energy (a).  The remaining turbulent energy that is
not collisionlessly damped at $ k_\perp \rho_i \sim 1$ is transferred
nonlinearly to a kinetic \Alfven wave cascade in the kinetic
dissipation range, $ k_\perp \rho_i \gtrsim 1$, leading to a magnetic
energy spectrum $E_B \propto k_\perp^{-2.8}$ (a) and a wavevector
anisotropy $k_\parallel \propto k_\perp ^{1/3}$ (b). In addition,
electron Landau damping becomes stronger as the wavenumber increases
over the entire range $ k_\perp \rho_i \gtrsim 1$ (a). Finally,
electron Landau damping becomes sufficiently strong to terminate the
kinetic turbulent cascade, leading to an exponential fall-off of the
magnetic energy spectrum at $ k_\perp \rho_e \sim 1$ (a), inhibiting
the transfer of energy to higher parallel wavenumber (b), and possibly
launching an electron entropy cascade (not shown).

\begin{acknowledgement}
I would like to thank Steve Cowley, Bill Dorland, and Eliot Quataert
for their invaluable guidance, steering my career toward the study of
turbulence in kinetic astrophysical plasmas.  Alex Schekochihin, Tomo
Tatsuno, Ryusuke Numata, and Jason TenBarge have contributed
tremendously to my efforts to understand kinetic turbulence.
Financial support has been provided by NSF grant PHY-10033446, NSF
CAREER Award AGS-1054061, and NASA grant NNX10AC91G. 
\end{acknowledgement}


\end{document}